\documentclass[aps,prb,twocolumn,groupedaddress]{revtex4}%
\usepackage{amsmath}
\usepackage{graphics}
\usepackage{graphicx}
\usepackage{epsfig}
\usepackage{amssymb}
\usepackage{amsfonts}%
\providecommand{\U}[1]{\protect\rule{.1in}{.1in}}

\begin{document}
\title{On the superconducting dome near antiferromagnetic quantum critical points}
\author{Mucio A. \surname{Continentino}}
\email{mucio@if.uff.br}
\affiliation{Instituto de F\'{\i}sica, Universidade Federal Fluminense, Campus da Praia
Vermelha, 24210-340, Niter\'oi, RJ, Brazil}
\date{\today }

\begin{abstract}
One of the most exciting discoveries in strongly correlated systems has been
the existence of a superconducting dome on heavy fermions close to the quantum
critical point where antiferromagnetic order disappears. It is hard even for
the most skeptical not to admit that the excitations which bind the electrons
in the Cooper pairs have a magnetic origin. As a system moves away from an
antiferromagnetic quantum critical point, (AFQCP) the correlation length of
the fluctuations decreases and the system goes into a local quantum critical
regime. The attractive interaction mediated by the non-local part of these
excitations vanishes and this allows to obtain an upper bound to the
superconducting dome around an AFQCP.

\end{abstract}
\maketitle


\section{Introduction}

The study of heavy fermion materials is an exciting area in physics motivating
sophisticated experimental work and giving rise to many new concepts and ideas
\cite{si}. In particular the fact that heavy fermions are close to a magnetic
quantum critical point \cite{base} has brought a new range of possibilities to
this field, both theoretically and experimentally.

On the course of their investigations, as experimentalists aim to reach closer
to the AFQCP at even lower temperatures, came out the exciting discovery of a
superconducting dome encircling a putative AFQCP \cite{lonzarich}. The region
of superconductivity in the phase diagram is restricted to a close
neighborhood of the AFQCP. Even for the most skeptical it is hard not to admit
that in this case superconductivity is due to quantum antiferromagnetic
fluctuations associated with the QCP \cite{nature}.

The theory of superconductivity mediated by spin fluctuations has progressed
very much in the last decades mostly due to its relevance for high temperature
superconductivity \cite{mon1,este,monden,monmag,km}. In these theories, the
paramagnon propagator describing critical antiferromagnetic fluctuations close
to an AFQCP can be written in the scaling form \cite{este},
\begin{equation}
\chi(q, \omega)=\frac{\chi_{S}}{i \omega\tau+ q^{2} \xi^{2} + 1} \label{qprop}%
\end{equation}
where $\chi_{S}=\chi_{0}/|g|$ is the staggered susceptibility, $\xi
=\sqrt{A/|g|}$ and $\tau=\tau_{0} \xi^{z}$ the correlation length and critical
relaxation time, respectively. The quantity $g$ measures the distance (in
energy scale) to the AFQCP (at $g=0$), $A$ is the stiffness of the spin
fluctuations, $\tau_{0}=1/A$ and the dynamic exponent $z=2$. We are interested
here in quantum phase transitions which occur in three ($d=3$) or two
dimensions ($d=2$). Since the dynamic exponent associated with an AFQCP takes
the value \cite{hertz} $z=2$, the effective dimension associated with the
antiferromagnetic quantum phase transition is $d_{eff}=d+2$. Then, for $2d$
and $3d$ systems the effective dimension coincides or is above the upper
critical dimension $d_{c}=4$, respectively. This implies that the transition
at $T=0$, $g=0$ is described by Gaussian or mean-field exponents \cite{livro}
with logarithmic corrections in the marginal case ($d=2$). Then for $d_{eff}
\ge d_{c}$ the Gaussian exponents, $\gamma=1$, for the staggered
susceptibility and $\nu=1/2$, for the correlation length turn out to describe
correctly the quantum critical behavior of the AFQCP.

The Gaussian free energy close to the AFQCP can be written as ($k_{B}=1$)
\cite{moriya,local},
\begin{equation}
f=-\frac{3}{\pi}\sum_{q} T \int_{0}^{\infty} \frac{d \lambda}{e^{\lambda}-1}
\tan^{-1}\left[  \frac{2 \pi\lambda T \xi^{z}}{A(1+q^{2} \xi^{2})}\right]
\label{free}%
\end{equation}
For temperatures $T << T_{coh}$ the free energy is given by,
\begin{equation}
f=-\frac{\pi^{2} T^{2} \xi^{z-d}}{A}\left(  \frac{L}{2 \pi}\right)  ^{d} S_{d}
\int_{0}^{q_{c} \xi} dy \frac{ y^{d-1}}{1+y^{2}}%
\end{equation}
where $q_{c}$ is a cut-off. The coherence temperature, $T_{coh}
=|g|^{\nu z}=|g|$ is that introduced by Continentino et al.
\cite{base} and marks the entrance of the system in the Fermi liquid
regime. $S_{d}$ is the surface of a d-dimensional sphere with unit
radius. The specific heat $C/T=-\partial^{2} f/\partial T^{2}$ in
the Fermi liquid regime, $T << T_{coh}$, is easily obtained,
\begin{equation}
C/T=\frac{V/\xi}{A} q_{c} \xi\left(  1-\frac{\tan^{-1}q_{c} \xi}{q_{c} \xi
}\right)
\end{equation}
in $3d$ and,
\begin{equation}
C/T=\frac{\pi S_{2}}{2A} \ln\left(  1+q_{c}^{2} \xi^{2} \right)
\end{equation}
in $2d$. In the critical regime $q_{c} \xi\gg1$, we get that
$\gamma=C/T$ is constant in $3d$ and logarithmically divergent in
$2d$ \cite{moriya}. We can also define \cite{lacroix,local} a
\emph{local limit}, $q_{c} \xi < 1$, in which case the specific heat
is given by,
\begin{equation}
C/T=\frac{2 \pi^{2} N}{T_{coh}}%
\end{equation}
independent of dimension. This result can be obtained directly from
Eq. \ref{free} neglecting its q-dependence and replacing $\sum_{q}
\rightarrow N$. The propagator associated with these  local spin
fluctuations is given by,
\begin{equation}
\chi_{L}(\omega)=\frac{\chi_{S}}{i \omega\tau+ 1} \label{lprop}%
\end{equation}
with $\chi_{S}=\chi_{0} / |g|$ and $\tau=\tau_{0} \xi^{z}$. It is
remarkable that in spite of the local character of the fluctuations
in this regime, the system is still \textit{aware} of the quantum
phase transition through the dependence of $\tau$ and $\chi_{S}$ on
$g$. Indeed, in this regime the fluctuations are local in space but
correlated along the \emph{ time directions}. The theory of this
regime can be described as a critical theory in $d_{eff}=z=2$ but
with Euclidean dimension $d=0$. The properties of the system for
$q_{c} \xi< 1$ have been described in Ref. \cite{local}. They can
all be expressed in terms of a single parameter, the coherence
temperature\cite{local}.

\section{Relation to superconductivity}

The q-dependent propagator given by Eq. \ref{qprop} provides an
attractive interaction in the singlet channel among quasi-particles
in neighboring sites. Thus, antiferromagnetic paramagnons can give
rise to d-wave pairing with $d_{x^{2}-y^{2}}$ symmetry \cite{este},
for $q_{c}\xi \gg 1$. What about local spin fluctuations?

The spatial dependence of the dynamic susceptibility can be obtained from the
Fourier transform of the $q$ and $\omega$ dependent propagator,
\begin{equation}
\chi(r,\omega)=\sum_{q}\chi(q,\omega)e^{iq.r}.
\end{equation}
In the case of local spin fluctuations, the relevant propagator is
given by Eq. \ref{lprop}, such that,
\begin{equation}
\Re e\chi_{L}(\omega)=\frac{\chi_{S}}{1+\omega^{2}\tau^{2}},
\end{equation}
and,
\[
\Re e\chi_{L}(\omega=0)=\chi_{S}.
\]
Then,
\[
\chi(r)=\chi_{S}\sum_{q}e^{iq.r}=2\pi\chi_{S}\int dqd\theta q^{2}\sin\theta
e^{iqr\cos\theta}.
\]

The interaction among the quasi-particles according to Monthoux et al.
\cite{nature} is given by:
\begin{align}
U(r) &  =-\lambda^{2}\chi(r)S_{1}.S_{2}\\
&  =-\lambda^{2}(-1)\frac{8\pi^{4}\chi_{S}}{a^{3}}\frac{1}{(\frac{\pi r}%
{a})^{3}}\left\{  \sin\frac{\pi r}{a}-\frac{\pi r}{a}\cos\frac{\pi r}%
{a}\right\}  .\nonumber
\end{align}
For small $r$ this yields,
\begin{align}
&  U(r\rightarrow0)=\allowbreak8\pi\frac{\lambda^{2}}{r^{3}}\chi_{S}\frac
{1}{3}\frac{\pi^{3}}{a^{3}}r^{3}=\allowbreak\frac{8}{3}\frac{\pi^{4}}{a^{3}%
}\lambda^{2}\chi_{S}\nonumber\\
&  U(a)=\allowbreak8\frac{\pi^{2}}{a^{3}}\lambda^{2}\chi_{S}\nonumber\\
&  U(2a)=\allowbreak-2\frac{\pi^{2}}{a^{3}}\lambda^{2}\chi_{S}\label{magcase}%
\end{align}
If we are dealing with a lattice then we have a delta-function at the origin.
The potential is repulsive at the origin and zero everywhere else in the
singlet channel.

Also, it is interesting to obtain the results in the case the local propagator
is associated with density fluctuations. In this case \cite{este},
\[
U(r)=-\lambda^{2}\chi(r)
\]
and we obtain,
\begin{align}
&  U(r\rightarrow0)=\allowbreak-8\pi\frac{\lambda^{2}}{r^{3}}\chi_{S}\frac
{1}{3}\frac{\pi^{3}}{a^{3}}r^{3}=\allowbreak-\frac{8}{3}\frac{\pi^{4}}{a^{3}%
}\lambda^{2}\chi_{S}\nonumber\\
&  U(a)=-\allowbreak8\frac{\pi^{2}}{a^{3}}\lambda^{2}\chi_{S}\nonumber\\
&  U(2a)=\allowbreak2\frac{\pi^{2}}{a^{3}}\lambda^{2}\chi_{S} \label{dencase}%
\end{align}
where $\chi_{S}$ in this case is the compressibility. In the case of a lattice
the interaction is attractive at the origin and zero everywhere else in the
singlet channel.

In the magnetic case, Eqs. \ref{magcase} show that the on-site and
nearest neighbor quasi-particle interaction mediated by critical
local spin fluctuations are repulsive in the singlet channel and do
not lead to Cooper pair formation. Then, as the system moves away
from the AFQCP, the correlation length of the spin fluctuations
decreases and for $q_{c} \xi \sim 1$ the relevant interactions
mediated by these fluctuations become repulsive everywhere
destroying superconductivity. On the other hand, local charge
fluctuations can still mediate an attractive local interaction.

The condition
\begin{equation}
q_{c} \xi=1 \label{yes}%
\end{equation}
puts an upper limit to the region of the phase diagram around the AFQCP where
superconductivity mediated by spin fluctuations can exist. At zero temperature
this implies that superconductivity survives up to a critical coupling
$|g|_{S}=Aq_{c}^{2}$.

In order to extend Eq. \ref{yes} to finite temperatures ($T$), we consider the
scaling form of the correlation length, $\xi= \sqrt{A}|g|^{-\nu}F[T/T_{coh}]$,
where $F[t]$ is a scaling function and $T_{coh}=|g|^{\nu z}$. In this case Eq.
\ref{yes} can be written as,
\begin{equation}
F\left[  \frac{T_{D}}{|g|^{\nu z}}\right]  =\frac{\sqrt{|g|}}{\sqrt{A} q_{c}}
\label{yes1}%
\end{equation}
where $T_{D}(g)$ represents an upper limit for superconductivity around the AFQCP.

\begin{figure}[tbh]
\centering{\includegraphics[scale=0.5,angle=270]{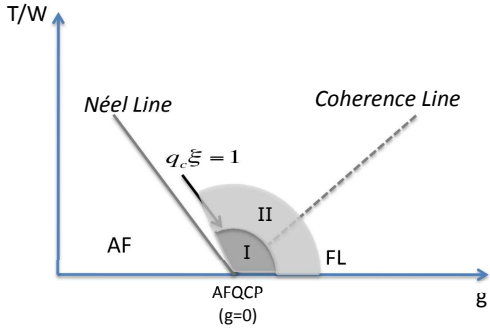}}\caption{(Color
online) The line {$q_{c} \xi(T_{D})=1$} represents an upper limit
where superconductivity induced by antiferromagnetic spin
fluctuations can occur. It's equation is {$T_{D}=|g|^{\nu
z}G^{-1}[|g|/Aq_{c}^{2}]$} (see equation \ref{yes1}) and is plotted
schematically in the figure. It separates region I with $q_{c}
\xi\gg1$ from region (II) of local quantum criticality where
fluctuations are critical in the time directions but local in space.
The latter gives rise to repulsive, on-site and nearest neighbors
interactions
(see text).}%
\label{fig1}%
\end{figure}

The scaling function $F(t)$ has a well known behavior in two
limiting cases. First, $F(t \rightarrow0) \propto1 - t^{2} + O\left(
t^{4}\right) $, such that $F(0)=1$ and for $t \ll 1$, i.e., $T \ll
T_{coh}$ this yields a Fermi liquid behavior for the staggered
susceptibility which in the spin fluctuation theory is used to
define the correlation length \cite{moriya}. Also, neglecting the
effect of dangerously irrelevant interactions (to be discussed
below), $F(t \rightarrow\infty) \propto t^{x}$ where the exponent
$x$ is determined by the condition that the dependence of the
correlation length on $g$ cancels out. This yields $x=-1/z$, such
that, at the quantum critical trajectory ($g=0$, $T \rightarrow0$),
the correlation length diverges as $\xi\propto T^{-1/z}$. We can
also show using this asymptotic behavior of the scaling function
that the point in the phase diagram at which the superconducting
temperature can attain its maximum value is just above the AFQCP,
i.e., at $g=0$. In fact $d T_{D}/d g \propto|g|^{z/2}$ and vanishes
at $g=0$.

An interpolation formula for the scaling function which gives correct results
on both limits ($t \rightarrow0$ and $\infty$) is given by,
\begin{equation}
F(t)=\frac{1}{(1+t^{2})^{1/4}},
\end{equation}
where we used the value of the exponents $\nu=1/2$ and $z=2$. Using this
expression for the scaling function, Eq. \ref{yes} can be written as,
\begin{equation}
\sqrt{T_{D}^{2} + |g|^{2}}=Aq_{c}^{2}.
\end{equation}
This has the form of a dome as shown in Fig. \ref{fig1}. This equation
provides a reasonable interpolation for the lines of constant correlation
length in the region of the phase diagram $g \ge0$. For negative ($g<0$),
there may be thermal fluctuations that change the scaling behavior of the
correlation length, as will be discussed below. The physical significance of
$T_{D}$ is now quite clear. It provides an upper limit to the region where
superconductivity induced by quantum antiferromagnetic spin fluctuations can
exist. While for $T \approx0$ this may provide a reasonable estimate of the
actual superconducting region and its shape, for larger temperatures thermal
fluctuations should reduce the critical temperature $T_{s}$ to values well
below $T_{D}$.

\begin{figure}[tb]
\centering{\includegraphics[scale=0.5]{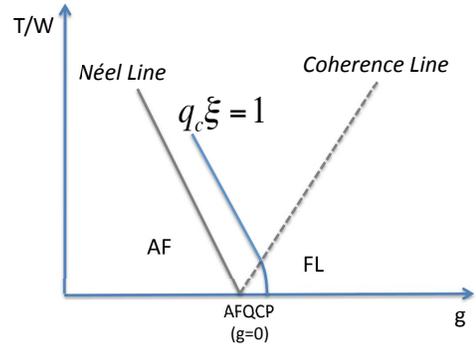}}\caption{(Color
online) Line of constant correlation length {$(q_{c} \xi=1)$} when
thermal or
interacting antiferromagnetic spin fluctuations are taken into account.}%
\label{fig2}%
\end{figure}

In the theory of the AFQCP for $d_{eff} \ge d_{c}$, the quartic
interaction $u$ is \textit{dangerously irrelevant} \cite{millis}. It
determines the shape of the Neel line and changes the value of the
shift exponent $\psi$ from the expected scaling result, $\psi=\nu
z$, to $\psi=z/(d+z-2)$ (for the AFQCP in 3d, $\psi=2/3$)
\cite{millis}. The scaling expression for the correlation length can
be immediately generalized to include the effect of the quartic
interaction $u$. It is given by \cite{livro}
\begin{equation}
\xi=\sqrt{A}|g(T)|^{-\nu}G\left[  \frac{T}{|g(T)|^{\nu z}} \right]
\label{general}%
\end{equation}
In this equation $g(T)=g-uT^{1/\psi}$, such that, $g(T_{N})=0$ gives the
equation for the N\'eel line. The scaling function $G(t)$ has the following
asymptotic behaviors; $G(t=0)=1$ to reproduce the previous zero temperature
results, and $G(t \rightarrow\infty) =t^{\frac{\tilde{\nu}-\nu}{\nu z}}$. The
latter guarantees the correct behavior close to the critical N\'eel line
$g(T_{N})=0$, i.e., $\xi=Q(T)|g(T))|^{-\tilde{\nu}}$, with the amplitude
$Q(T)=\sqrt{A}T^{\frac{\tilde{\nu}-\nu}{\nu z}}$. When $T \rightarrow T_{N}$,
$g(T) \rightarrow0$ and the correlation length diverges with the thermal
correlation length exponent $\tilde{\nu}$. Assuming \cite{millis}, $\tilde
{\nu}=\nu=1/2$, we get at $g=0$,
\begin{equation}
\xi=\sqrt{\frac{A}{u}}T^{-1/3} \label{xigen}%
\end{equation}
For $g=0$, this yields a temperature for the dome at $g=0$,
$T^{u}_{D}(g=0)=\left(  Aq_{c}^{2}/u\right)  ^{3/2}$, to be compared
with
$T_{D}(g=0)=Aq_{c}^{2}$, obtained previously. This leads to $T_{D}=u(T_{D}%
^{u})^{2/3}$ and since $u$ is a small number, we expect that in
general $T_{D} \ll T_{D}^{u}$, such that, non-Gaussian fluctuations
allow in principle for larger critical superconducting temperatures
just above the AFQCP. However, in this case the lines of constant
$\xi$ satisfying Eq. \ref{yes} should follow closely the N\'eel line
as in Fig.\ref{fig2}. The experimental results show however that the
superconducting region has a dome shape \cite{lonzarich}. Then, they
seem to imply that only purely Gaussian quantum fluctuations are
effective in pairing the quasi-particles. Also notice that along the
N\'eel line, for $T \ne 0$, the quartic interaction is a
\textit{relevant} interaction \cite{livro}. Then, as the spin
fluctuations start to interact they apparently loose their efficacy
in pairing the quasi-particles. At least this is what the
experiments seem to imply.

In some heavy fermion systems as one moves away from the AFQCP, for example,
applying pressure in the system there is a second superconducting dome
\cite{flouquet,miyake}. This is generally attributed to pairing due to charge
fluctuations associated with a valence transition \cite{miyake}. This second
dome is larger than that associated with the AFQCP extending in a wider region
of pressures and temperature. As pointed out before in the case of pairing by
charge fluctuations the interaction is attractive when the system is in the
regime of local quantum criticality. So, we expected that in this case,
superconductivity can occur in a larger region of the phase diagram around the
relevant quantum critical point, as is in fact observed.

The energy scale of the \emph{magnetic glue} that fixes the region
where superconductivity can exist is given by $Aq_{c}^{2}$. $A$ is
the stiffness of the spin fluctuations and $q_{c}$ a cut-off
appropriate for a hydrodynamic description of these modes. This
quantity plays a role similar to the Debye energy in BCS
superconductors.

The region of the phase diagram just above the limiting
superconducting dome is a state of \textit{local quantum
criticality}. This state is characterized by a single energy scale,
the coherence temperature, $T_{coh}=|g|^{\nu z}$. It has a
resistivity which scales as $\rho\propto(T/T_{coh})^{2}$ for $T \ll
T_{coh}$ and as $\rho\propto(T/T_{coh})$ for $T \gg T_{coh}$. In the
case of anisotropic lattices, the crossover to the local regime
occurs in stages. For a tetragonal system with a spectrum of spin
fluctuations given by, $a_{xy} q_{x}^{2} + a_{xy} q_{y}^{2} +
a_{z}q_{z}^{2}$, as the system moves away from the AFQCP it goes
from $d=3$ to $d=2$ and finally to $d=0$ quantum critical behavior.

The dome shape of the superconducting region seems to indicate that the
interaction between the spin fluctuations acts in detriment of
superconductivity. It can not be excluded that interacting spin fluctuations
can still provide a pairing mechanism to produce, for example, a pseudo-gap
state, but not superconductivity. We have shown quite generally that if
Gaussian quantum spin fluctuations give rise to superconductivity, the maximum
allowed $T_{c}$ can be found just above the AFQCP. Finally, our results
strongly support the proposal that the second superconducting dome observed in
some heavy fermions systems is due to pairing by charge fluctuations. Since
even in the local quantum regime, charge fluctuations give rise to attractive
interactions, superconductivity in this case can extend over a wider region of
the phase diagram.

Our results are appropriate to describe the system in the paramagnetic region.
In the long range ordered magnetic phase of the diagram there are new
excitations, the spin waves, associated with transverse modes. Then in this
region the present approach does not provide any insight.

\acknowledgements I would like to thank Dr. P. Monthoux for
invaluable discussions. This work is partially supported by the
Brazilian agencies CNPq and FAPERJ.


\begin{thebibliography}{99}
\bibitem {pau}In the original spin fluctuation theory of the QCP
\cite{moriya}, the coherence temperature is never explicitly introduced.
However, it is clearly there, with the same significance that it was first
introduced in Ref. \cite{base}, as marking the onset of Fermi liquid behavior.
\end{thebibliography}
\end{document}